\def\bar{\begin{eqnarray}}
\def\ear{\end{eqnarray}}
\def\bb{\bibitem}
\def\eqi{\begin{equation}}
\def\eqf{\end{equation}}
\def\eqia{\begin{eqnarray}}
\def\eqfa{\end{eqnarray}}

\def\oc2{$\mathcal{O}(c^{-2})$}

\documentclass[11pt]{article}
\usepackage{amsmath,amsthm,amscd,amssymb}
\usepackage{latexsym}
\usepackage{graphicx,epsfig}

\begin{document}

\noindent{\bf \LARGE{Reply to ``The meaning of systematic errors, a comment to $^{\backprime\backprime}$Reply to On the Systematic Errors in the Detection of the Lense-Thirring Effect with a Mars Orbiter$^{\prime\prime}$, by Lorenzo Iorio'', by G. Felici}}
\\
\\
\\
{Lorenzo Iorio}\\
{\it Viale Unit$\grave{a}$ di Italia 68, 70125\\Bari (BA), Italy
\\tel. 0039 328 6128815
\\e-mail: lorenzo.iorio@libero.it}

\begin{abstract}
In this note we reply to the criticisms by Felici  concerning some
aspects of the recent frame-dragging test performed by Iorio with
the Mars Global Surveyor (MGS) spacecraft in the gravitational
field of Mars.
\end{abstract}



\begin{itemize}

\item
It seems that Felici (2007) argues that one of the ``elementary mistakes" by Iorio (2007a) would be that
the 1.6 meters residuals refer to a period of a few days
only and not to one year. Apart from the fact that the time span $\Delta P$ chosen by Iorio (2007a) does not amount
to one year and that Felici (2007) does not yield any explicit proofs or, at least, reasonings supporting his claim, if it was right the time series of the MGS normal orbit overlap differences used in (Iorio 2007a) should not exhibit any well-defined time-dependent pattern. It is not so: as shown in (Iorio 2007a), a linear trend is, in fact, present in agreement with the predictions of general relativity.
\item
Referring to Iorio (2007b), Felici (2007) writes: ``Here Iorio is writing: $^{\backprime\backprime}$the systematic errors calculated by Sindoni
et al. are not detected in the residuals of MGS, thus they are null.$^{\prime\prime}$". It is false\footnote{This error is amended in the v2 version of (Felici 2007).}: such a  statement is not present in (Iorio 2007b).
\item
In the present context with systematic errors we mean the possible action of all the unmodelled/mismodelled
forces acting on MGS. For example, from this point of view, also the Lense-Thirring effect itself can be regarded as a systematic bias because the gravitomagnetic force was not modelled at all in the orbit data reduction softwares used. As pointed out in several places, the RMS orbit overlap differences are just used to account, in general, for all the measuremnt/systematic errors giving an indication of the total orbit accuracy (Montenbruck and Gill 2000, Tapley et al. 2004, Lucchesi and Balmino 2006). The important point is that they cancel out, by construction, errors, systematic or not, common to consecutive arcs$-$it would just be the case of a bias like that described by Felici (2007)$-$, while effects like the Lense-Thirring one, accumulating in time, are, instead, singled out (Lucchesi and Balmino 2006).

Again, more or less sound speculations about the nature and the origin of the data points used in (Iorio 2007a) must ultimately cope with the real world. Two independent tests were performed so far supporting the Lense-Thirring hypothesis: the average over $\Delta P$ of the time series of the MGS out-of-plane RMS overlap differences  and its straight-line fit.
\end{itemize}

\section*{Acknowledgments}
I gratefully thank D.M. Lucchesi for interesting discussions about the orbit overlap differences.


\end{document}